\documentclass{llncs}[11pt]
\usepackage{url}
\usepackage{latexsym}

\newcommand{\triple}[3]{\langle#1,#2,#3\rangle}

\newcommand{\theString}{\tau}
\newcommand{\numOfRuns}[1]{\textit{run}(#1)}
\newcommand{\previousBound}{\frac{3}{1+\sqrt{5}}}

\newcommand{\theStringLength}{60064}
\newcommand{\theStringRuns}{56714}
\newcommand{\theStringTwoRuns}{113448}
\newcommand{\theStringThreeRuns}{170181}
\newcommand{\theStringThreeTwoDiff}{56733}
\newcommand{\theStringConstDiff}{18}
\newcommand{\newsingleFracValue}{0.944226}
\newcommand{\newBoundApprox}{0.944542}
\newcommand{\newsingleFrac}{\theStringRuns/\theStringLength}

\newcommand{\newBound}{\theStringThreeTwoDiff/\theStringLength}
\newcommand{\newBoundFrac}{\frac{\theStringThreeTwoDiff}{\theStringLength}}
\newcommand{\newBoundCalc}{\frac{\theStringThreeRuns - \theStringTwoRuns}{\theStringLength}}

\title{New Lower Bounds for the Maximum Number of Runs in a String}

\author{
Kazuhiko Kusano\inst{1}
\and 
Wataru Matsubara\inst{1}
\and
Akira Ishino\inst{1}
\and
Hideo Bannai\inst{2}
\and
Ayumi Shinohara\inst{1}
}
\institute{
Graduate School of Information Science, Tohoku University, \\
Aramaki aza Aoba 6-6-05, Aoba-ku, Sendai 980-8579, Japan \\
\email{\{kusano@shino., matsubara@shino., ishino@, ayumi@ \}ecei.tohoku.ac.jp}
\and
Department of Informatics, Kyushu University,\\
744 Motooka, Nishiku, Fukuoka 819-0395 Japan.\\
\email{bannai@i.kyushu-u.ac.jp}
}

\begin{document}
\maketitle

\begin{abstract}
 We show a new lower bound for the maximum number of runs in a string.
We prove that for any $\varepsilon > 0$,
 $(\alpha-\varepsilon)n$ is an asymptotic lower bound, where
 $\alpha=\newBound \approx \newBoundApprox$.
 It is superior to the previous bound $3/(1+\sqrt{5}) \approx 0.927$
 given by Fran\v{e}k {\em et  al.}~\cite{franek03:_maxim_number_of_runs_in_strin,franek06:_asymp_lower_bound_for_maxim}.
Moreover, our construction of the strings and the proof is much simpler than theirs.
\end{abstract}


\section{Introduction}
Repetitions in strings is an important element in the analysis and
processing of strings.
It was shown in~\cite{kolpakov99:_findin_maxim_repet_in_word} that
when considering {\em maximal repetitions}, or {\em runs}, 
the maximum number of runs $\rho(n)$ in any string of length $n$
is $O(n)$, leading to a linear time algorithm for
computing all the runs in a string.
Although they were not able to give bounds for the constant factor,
there have been several works to this end~\cite{rytter06:_number_of_runs_in_strin,rytter07:_number_of_runs_in_strin,crochemore07:_maxim_repet_in_strin}.
The currently known best upper bound\footnote{Presented on the website \url{http://www.csd.uwo.ca/faculty/ilie/runs.html}} is $\rho(n) \leq 1.048n$,
obtained by calculations based on the proof technique 
of~\cite{crochemore07:_maxim_repet_in_strin}.
The technique bounds the number of runs for each string by considering runs
in two parts: runs with long periods, and runs with short periods.
The former is more sparse and easier to bound while the latter is 
bounded by an exhaustive calculation concerning
how runs of different periods can overlap in an interval 
of some length.
On the other hand, an asymptotic lower bound on $\rho(n)$ is
presented in~\cite{franek06:_asymp_lower_bound_for_maxim},
where it is shown that for any $\varepsilon>0$, 
there exists an integer $N>0$ such that for any $n>N$,
$\rho(n) \geq (\alpha - \varepsilon)n$,
where
$\alpha= \previousBound \approx 0.927$.
It was conjectured in~\cite{franek03:_maxim_number_of_runs_in_strin}
that this bound is optimal.

In this paper, we prove that the conjecture was false, by showing a
new lower bound $\alpha=\newBound \approx \newBoundApprox$.
First we show a concrete string $\theString$ of length $\theStringLength$, which
contains $\theStringRuns$ runs in it.
It immediately disproves the conjecture, since $\newsingleFrac \approx \newsingleFracValue$
is already higher than the previous bound $0.927$.
Then we prove that the string $\theString^{k}$, which is the
string obtained by concatenating $k$ copies of $\theString$,
contains $\theStringThreeTwoDiff k-\theStringConstDiff$ runs for
any $k \geq 2$.
Since $|\theString^{k}| = \theStringLength k$, it yields the new lower bound
$\newBound$ as $k \rightarrow \infty$.

%



\section{Preliminaries}

Let $\Sigma$ be a finite set of symbols, called an {\em alphabet}. 
Strings $x$, $y$ and $z$ are said to be a \emph{prefix},
\emph{substring}, and \emph{suffix} of the string $w=xyz$, respectively.
The length of a string $w$ is denoted by $|w|$. 
The $i$-th symbol of a string $w$ is denoted by $w[i]$ for $1 \leq i \leq |w|$, and 
the substring of $w$ that begins at position $i$ and ends
at position $j$ is denoted by $w[i:j]$ for $1\leq i \leq j \leq |w|$.
A string $w$ has period $p$ if $w[i] = w[i+p]$ for $1\leq i\leq |w|-p$.
A string $w$ is called {\em primitive} if $w$ cannot be written as $u^k$, 
where $k$ is a positive integer, $k \geq 2$.

A string $u$ is a {\em run} if it is periodic with (minimum) period $p \leq
|u|/2$. 
A substring $u=w[i:j]$ of $w$ is a {\em run in $w$} if it is a run of
period $p$ and neither $w[i-1:j]$ nor $w[i:j+1]$ is a run of period $p$,
that means the run is maximal.
We denote the run $u=w[i:j]$ in $w$ by the triple
$\triple{i}{j\!-\!i\!+\!1}{p}$ consisting of the begin position
$i$, the length $|u|$, and the minimum period $p$ of $u$.
A run of $w$ which is a prefix (resp. suffix) of $w$ is called a prefix
(resp. suffix) run of $w$,
For a string $w$, we denote by $\numOfRuns{w}$ the number of runs in
$w$. 

For example, the string {\tt aabaabaaaacaacac} contains the following 7 runs:
$\triple{1}{2}{1} = \texttt{a}^2$, 
$\triple{4}{2}{1} = \texttt{a}^2$, 
$\triple{7}{4}{1} = \texttt{a}^4$,
$\triple{12}{2}{1} = \texttt{a}^2$,
$\triple{13}{4}{2} = \texttt{(ac)}^2$, 
$\triple{1}{8}{3} = \texttt{(aab)}^{\frac{8}{3}}$, and 
$\triple{9}{7}{3} = \texttt{(aac)}^{\frac{7}{3}}$.
%
Thus $\numOfRuns{\tt aabaabaaaacaacac} = 7$.

We are interested in the behavior of the {\em maxrun function} defined by
\[
 \rho(n) = \max \{ \numOfRuns{w} \mid w \mbox{ is a string of length $n$} \}.
\]

Fran\v{e}k, Simpson and
Smyth~\cite{franek03:_maxim_number_of_runs_in_strin} showed a beautiful
construction of a series of strings which contains many runs, and later
Fran\v{e}k and Qian Yang~\cite{franek06:_asymp_lower_bound_for_maxim}
formally proved a family of true asymptotic 
lower bounds arbitrarily close to $\previousBound n$ as follows.
\begin{theorem}[\cite{franek06:_asymp_lower_bound_for_maxim}]
For any $\varepsilon > 0$ there exists a positive integer $N$ so that
 $\rho(n) \geq \left(\previousBound - \varepsilon\right)n$ for
 any $n \geq N$.
\end{theorem}



\section{Basic Properties}

In this section, we summarize some basic properties concerning periods
and repetitions in strings, which will be utilized in the sequel.

The next Lemma given by Fine and Wilf~\cite{fine65:_uniqueness_theorems_for_periodic_functions} provides an important property on
periods of a string. 
\begin{lemma}[Periodicity Lemma~(see 
\cite{lothaire02:_algebraic_combinatorics_on_words,crochemore02:_jewels_of_stringology})]
\label{lemma:periodicity lemma}
Let $p$ and $q$ be two periods of a string $w$.
If $p+q-\gcd(p,q)\le |w|$, then $\gcd(p,q)$ is also a period of $w$.
\end{lemma}

For a string $w$, let us consider a series of strings $w$, $w^2$, $w^3$,
$w^4 \ldots$, and observe all runs contained in these strings.
There are many cases, which confuse the task of counting the number of runs in
these strings.
\begin{enumerate}
 \item A run in $w^{k}$ which is neither a suffix nor prefix run
   of $w^{k}$
   is also a run in $w^{k+1}$.
 \item A suffix run in $w^{k}$ and a prefix run in $w$ may be merged
       into one run in $w^{k+1}$.
 \item A suffix run in $w^{k}$ may be extended to a run in $w^{k+1}$.
 \item A new run may be newly created at the border between $w^{k+1}$
       and $w$.
\end{enumerate}

Concerning case 4, note that a new run that did not appear in
$w$ or $w^2$ may be created in $w^3$.
For example, consider strings $w=\texttt{abcacabc}$, and 
$r = (\texttt{cabca})^2$.
We can verify that $r$ is a run $\triple{8}{10}{5}$ of $w^3 =
\texttt{abcacab\underline{cabcacabca}bcacabc}$, while $r$ does not
appear in $w^2 = \texttt{abcacabcabcacabc}$.
Moreover, the same argument holds also for binary alphabet
${\texttt{0},\texttt{1}}$; Replace \texttt{a}, \texttt{b}, \texttt{c}
into \texttt{01}, \texttt{10}, \texttt{00}, respectively in the above example.

However, the following lemma shows that the length of such new runs
can be bounded.



\begin{lemma}
\label{lemma: l is less than 2n}
Let $w$ be a string of length $n$.
For any $k \geq 3$, let $r = \triple{i}{l}{p}$ be a run in $w^{k}$.
If $l \geq 2n$, then $i=1$ and $l=kn$, that is, $r = w^k$.
\end{lemma}

\begin{proof}
We assume that $n>1$, since it is trivial for the case $n=1$.
Since $p$ is the minimum period of the run $r$, we know 
$|r| = l \geq 2p$ and $l \geq 2n$ . 
Let $u$ be a primitive string of length $m$ where $w=u^t$
for some integer $t \geq 1$.
Then, $|u| = m \leq n$ is also a period of run $r$.
Since $p + m \leq l$ , Lemma~\ref{lemma:periodicity lemma}
 claims that $\gcd(p,m)$ is also a period of run $r$.
If $p > m$, then $\gcd(p,m) < p$, which contradicts the assumption that 
$p$ is the minimum period of $r$.
If $p < m$, then 
it contradicts the assumption that $u$ is primitive.
Therefore we have $p=m$.
Since $m$ is a period of $w^k$, we have $r=\triple{1}{kn}{m}=w^k$.
\end{proof} 

This lets us prove the following lemma which gives a formula for $\numOfRuns{w^k}$.
\begin{lemma}
\label{lemma: general term of run(w^k)}
Let $w$ be a string of length $n$.
For any $k \geq 2$, 
$\numOfRuns{w^k} = Ak - B$, 
where $A = \numOfRuns{w^3} - \numOfRuns{w^2}$ and
 $B = 2\numOfRuns{w^3} - 3\numOfRuns{w^2}$.
\end{lemma}

\begin{proof}
We think about the increase in the number of runs,
when concatenating $w^k$ and $w$.
Let $r = \triple{i}{l}{p}$ be a run of $w^{k+1}$
such that $i+l > nk+1$, that is, $r$ ends somewhere in the last $w$ of $w^{k+1}$.
By Lemma~\ref{lemma: l is less than 2n}, if $i \leq (k-2)n$ then
$r = w^{k+1}$.
In such a case, $r$ does not increase the number of runs
since the run will have already been considered in $w^2$.
Therefore, the increase in runs can be considered by restricting our
attention to runs with $i > (k-2)n$, that is, 
the increase in runs for the last 3 $w$'s of $w^{k+1}$
when concatenating $w$ to the last 2 $w$'s of $w^{k}$.
This gives us $\numOfRuns{w^{k+1}} - \numOfRuns{w^k} = \numOfRuns{w^3} - \numOfRuns{w^2}$.





\begin{eqnarray*}
	\numOfRuns{w^k} & = & \numOfRuns{w^{k-1}} + \numOfRuns{w^3} - \numOfRuns{w^2} \\
	                & = & \numOfRuns{w^{k-2}} + 2(\numOfRuns{w^3} - \numOfRuns{w^2}) \\
	                & = & \numOfRuns{w^2} + (k-2)(\numOfRuns{w^3} - \numOfRuns{w^2}) \\
	                & = & k(\numOfRuns{w^3} - \numOfRuns{w^2}) - (2\numOfRuns{w^3} - 3\numOfRuns{w^2}) 
\end{eqnarray*}
for $k \geq 3$. It is easy to see that the equation also holds for $k=2$.
\end{proof}

\begin{theorem}
\label{theorem:general_bound_theorem}
For any string $w$ and any $\varepsilon > 0$, there exists a positive integer $N$ such that
 for any $n \geq N$,
\[
 \frac{\rho(n)}{n} > \frac{\numOfRuns{w^3} - \numOfRuns{w^2}}{|w|} - \varepsilon.
\]
\end{theorem}

\begin{proof}
By Lemma~\ref{lemma: general term of run(w^k)},
$\numOfRuns{w^k} = Ak - B$, 
where $A = \numOfRuns{w^3} - \numOfRuns{w^2}$ and
 $B = 2\numOfRuns{w^3} - 3\numOfRuns{w^2}$.

For any given $\varepsilon > 0$, we choose $N >
 \frac{A-B}{\varepsilon}$.
For any $n \geq N$, let $k$ be the integer satisfying
$|w|(k-1) \leq n < |w| k$.
Notice that $k > \frac{n}{|w|} \geq \frac{N}{|w|} \geq \frac{A-B}{|w| \varepsilon}$.
Since $\rho(i+1) \geq \rho(i)$ for any $i$, and 
 $|w^{k-1}| = |w|(k-1)$, 
\begin{eqnarray*}
 \frac{\rho(n)}{n}  & \geq &
 \frac{\rho(|w|(k-1))}{|w|k} 
 \geq \frac{\numOfRuns{w^{k-1}}}{|w|k} 
  =  \frac{A(k-1)-B}{|w|k}
  =  \frac{Ak-A-B}{|w|k}\\
  & =  &\frac{A}{|w|}-\frac{A-B}{|w|k}
   > \frac{A}{|w|}- \varepsilon.
\end{eqnarray*}
\qed
\end{proof}



\section{New Lower Bounds}

We found some strings which contain many runs, by running a computer
program which utilizes a simple heuristic search for run-rich binary
strings.
Given a buffer size, the search
first starts with the single string \texttt{0} in the buffer.
At each round, two new strings are created from each string
in the buffer by appending \texttt{0} or \texttt{1} to the string.
The new strings are then sorted in order of
$\numOfRuns{w^3} - \numOfRuns{w^2}$, and only those that fit in the
buffer are retained for the next round.
Strings that give a high ratio of runs are recorded.

We tried several variations of the algorithm, and found many run-rich strings.
Among these strings found so far, the string $\theString$, 
lets us prove the currently best lower bound on the maximum number 
of runs in a string.
Since $\theString$ is too long to include in the paper,
we will make $\theString$ available on our web site
\footnote{\url{http://www.shino.ecei.tohoku.ac.jp/runs/}}.
Once we have $\theString$, it is straightforward to
confirm that the following lemma holds.
Any na\"{i}ve program to count runs in a string would be sufficient.

\begin{lemma}
\label{lemma:best_runrich_string}
There exists a string $\theString$ such that
$|\theString| = \theStringLength$,
$\numOfRuns{\theString}=\theStringRuns$,
$\numOfRuns{\theString^2}=\theStringTwoRuns$,
and $\numOfRuns{\theString^3}=\theStringThreeRuns$.
\end{lemma}

It immediately disproves the conjecture, since $\newsingleFrac \approx \newsingleFracValue$
is already higher than the previous bound $\previousBound \approx 0.927$.
We now show the main result of this paper.
\begin{theorem}
For any $\varepsilon > 0$ there exists a positive integer $N$ so that\\
 $\rho(n) > \left(\alpha - \varepsilon\right)n$ for
 any $n \geq N$, where $\alpha = \newBoundFrac\approx\newBoundApprox$.
\end{theorem}

\begin{proof}
  From Theorem~\ref{theorem:general_bound_theorem} and
  Lemma~\ref{lemma:best_runrich_string}, we have
\[
\frac{\rho(n)}{n} > 
\newBoundCalc - \varepsilon =
\newBoundFrac - \varepsilon.
\]
\qed
\end{proof}

For proof of concept, we present in the Appendix,
a shorter string $\theString_{1558}$ with
$|\theString_{1558}| = 1558,
\numOfRuns{\theString_{1558}} = 1445,
\numOfRuns{\theString_{1558}^2} = 2915,
\numOfRuns{\theString_{1558}^3} = 4374$
that gives a smaller bound 
$(4374-2915)/1558 \approx 0.93645$
compared to $\theString$, 
but is still better than previously known.



\section{Conclusion}
We presented a new lower bound $\newBound \approx \newBoundApprox$ for the maximum
number of runs in a string. The proof was very simple, once after we
verified that the runs in the string $\theString$ is $\theStringRuns$, and 
noticed some trivial properties of the string.
We do not think that the bound is optimal.
We believe that our work would revive the interests to push the lower
bound higher up, since the previous bound $3/(1+\sqrt{5}) \approx 0.927$
was conjectured to be the optimal since 2003.


\bibliographystyle{splncs}


\newpage
\section*{Appendix}

The binary string $\theString_{1558}$
with
$|\theString_{1558}| = 1558,
\numOfRuns{\theString_{1558}} = 1445,
\numOfRuns{\theString_{1558}^2} = 2915,
\numOfRuns{\theString_{1558}^3} = 4374$,
giving lower bound $(4374-2915)/1558 \approx 0.93645 > 0.927$.\\

\noindent
\texttt{%
110101101001011010110100101101011001101011010010110101101001011010\\
110010110101101001011010110100101101011001101011010010110101101001\\
011010110010110101101001011010110010110100101101011010010110101100\\
101101011010010110101101001011010110010110100101101011010010110101\\
100101101011010010110101100101101001011010110100101101011001011010\\
110100101101011010010110101100101101011010010110101100101101001011\\
010110100101101011001011010110100101101011010010110101100101101001\\
011010110100101101011001011010110100101101011001011010010110101101\\
001011010110010110101101001011010110100101101011001011010110100101\\
101011001011010010110101101001011010110010110101101001011010110010\\
110100101101011010010110101100101101011010010110101101001011010110\\
010110100101101011010010110101100101101011010010110101100101101001\\
011010110100101101011001011010110100101101011010010110101100101101\\
011010010110101100101101001011010110100101101011001011010110100101\\
101011010010110101100101101001011010110100101101011001011010110100\\
101101011001011010010110101101001011010110010110101101001011010110\\
100101101011001011010110100101101011001011010010110101101001011010\\
110010110101101001011010110010110100101101011010010110101100101101\\
011010010110101101001011010110010110100101101011010010110101100101\\
101011010010110101100101101001011010110100101101011001011010110100\\
101101011010010110101100101101011010010110101100101101001011010110\\
100101101011001011010110100101101011010010110101100101101001011010\\
110100101101011001011010110100101101011001011010010110101101001011\\
0101100101101011010010110101101001011010}


\vspace{5mm}
By interpreting $\theString_{1558}$ as a binary representation of an integer, 
it can be expressed in hexagonal representation by:\\

\noindent
\texttt{%
0x35A5AD2D66B4B5A5ACB5A5AD2D66B4B5A5ACB5A5ACB4B5A5ACB5A5AD2D65A5AD\\
2D65AD2D65A5AD2D65AD2D696B2D696B2D2D696B2D696B4B59696B4B596B4B5969\\
6B4B596B4B5A5ACB5A5ACB4B5A5ACB5A5ACB4B5A5ACB5A5AD2D65A5AD2D65AD2D6\\
5A5AD2D65AD2D696B2D696B2D2D696B2D696B4B59696B4B596B4B59696B4B596B4\\
B5A5ACB5A5ACB4B5A5ACB5A5ACB4B5A5ACB5A5AD2D65A5AD2D65AD2D65A5AD2D65\\
AD2D696B2D696B2D2D696B2D696B4B59696B4B596B4B59696B4B596B4B5A5A}

\end{document}